\documentclass[aps,prc,twocolumn,nofootinbib,showpacs,floatfix]{revtex4-1}

\usepackage{graphicx}
\usepackage{amssymb}
\usepackage{amsmath}
\usepackage{bm}
\usepackage[usenames]{color}
\usepackage{ulem}
\usepackage{tensor}

\begin{document}

\title[]{Production cross section of neutron-rich isotopes with radioactive and stable beams}

\author{Myeong-Hwan \surname{Mun}}
\email{mhmun@ibs.re.kr}
\affiliation{Department of Physics, Kyungpook National University, Daegu, 702-701, Korea}
\affiliation{Rare Isotope Science Project, Institute for Basic Science, Daejeon, 305-811, Korea}

\author{G. G. \surname{Adamian} and N. V. \surname{Antonenko}}
\affiliation{Joint Institute for Nuclear Research, 141980 Dubna, Russia}

\author{Yongseok \surname{Oh}}
\email{yohphy@knu.ac.kr}
\affiliation{Department of Physics, Kyungpook National University, Daegu, 702-701, Korea}
\affiliation{Asia Pacific Center for Theoretical Physics, Pohang, Gyeongbuk 790-784, Korea}

\author{Youngman \surname{Kim}}
\affiliation{Rare Isotope Science Project, Institute for Basic Science, Daejeon, 305-811, Korea}


\begin{abstract}

The production cross section of neutron-rich isotopes of \nuclide[][]{Ca},
\nuclide[][]{Zn}, \nuclide[][]{Te}, \nuclide[][]{Xe}, and \nuclide[][]{Pt} are
predicted in the diffusive multi-nucleon transfer reactions with stable and
radioactive beams.
With these isotopes one can treat the neutron shell evolution beyond $N=28$,
$50$, $82$, and $126$.
Because of the small cross sections, the production of nuclei near the neutron
drip line requires the optimal choice of reaction partners and bombarding
energies.
\end{abstract}



\maketitle

\section{INTRODUCTION}

One of the primary issues of nuclear physics is an expansion of the present limits
of the nuclear chart far from the line of stability.
The neutron-rich isotopes located close to the neutron-drip line
have attracted interests during the last few decades.
In the study of exotic neuron-rich nuclei a variety of new aspects have been
discovered such as the existence of neutron skins, neutron halos, and shell
evolutions leading to new magic numbers.
Furthermore, these neutron-rich nuclei play a crucial role in the $r$-process
in nuclear synthesis.

Nowadays, the nuclei far from stability are accessible in vast region of the
nuclear chart by various reactions.
Fission should be suitable to generate a certain number of neutron-rich
nuclides~\cite{HM00}, but this method was found not to be promising for the
production of nuclei near the drip line.
Instead, the fragmentation reactions can produce very neutron-rich
nuclei~\cite{BSCEF99-BFAB08,KAGP11} at energies well above the Fermi energy.
However, in the fragmentation reactions, the yields of these nuclei decrease
with increasing isospin quantum number and a large excitation energy is
available, which can wash out the primary yields of extremely neutron-rich nuclei.
On the other hand, the neutron-rich nuclei can be also effectively produced in
the transfer-type reactions in which the control of excitation energy is
easier~\cite{Volkov78,SH85,Volkov89,VBGK87-KVBG90-SVSW91,BQDJ90,ANAJ94,VV01,%
FZJH04-KBFP10,Lunardi09,RBNM07,CPS09,PAA05-PAA06,AAL10,KAAS10-KAAS11-KAASW11}.

The experimental setups, which are presently under construction, would provide
quite intensive beams of radioactive nuclei. (See, for example, Ref.~\cite{KoRIA13}.)
One of the applications of these beams would be the production of new neutron-rich
isotopes using either complete fusion or transfer reactions.
While one can reach only specific isotopes in the complete fusion reactions,
the transfer-type reactions can yield larger variety of isotopes because of the
restrictions in the choice of the target.
Note that the use of the beam of neutron-rich nuclei in the fragmentation
reactions has no advantage compared with the beam of stable isotopes.

The aim of the present paper is to predict the production cross sections of new
neutron-rich \nuclide[][]{Ca}, \nuclide[][]{Zn}, \nuclide[][]{Te}, \nuclide[][]{Xe},
and \nuclide[][]{Pt} isotopes via the diffusive multi-nucleon transfer reactions
at incident energies near the corresponding Coulomb barriers (3--5 MeV/nucleon).
These very neutron-rich isotopes have the neutron numbers beyond $N=28$, $50$,
$82$, and $126$, respectively.
So, the study on their productions is expected to provide useful information
about the shell evolution near the drip line.
Because the cross sections for producing neutron-rich isotopes near the neutron
drip line are estimated to be small, the production of neutron-rich isotopes
requires an optimal choice of the projectile-target combinations and bombarding
energies for future experiments.
In the present paper, by extending the work of Ref.~\cite{AAL10}, we investigate
the production of various neutron-rich isotopes and study other possible reactions
for the production of the isotopes previously considered.

The present paper is organized as follows.
In Sec.~\ref{sec:model}, we present the model used for estimating the production
cross sections of neutron-rich nuclei.
The obtained results are given in Sec.~\ref{sec:results} with detailed discussions.
We then summarize in Sec.~\ref{sec:summary}.

\section{Model} \label{sec:model}

The dinuclear system (DNS) concept~\cite{Volkov86} was proposed to explain the
mechanisms of the complete fusion, quasifission, and nucleon transfer reactions.
The quasifission and transfer products result from the decays of various DNS
configurations.
The DNS is formed at the capture stage of the reaction after complete dissipation
of the kinetic energy of the relative
motion~\cite{PAA05-PAA06,AAL10,ACNPV93-ACNPV95,AAS97-AASV97a-AASV97b,AAS99,%
AAS03,AASS10}.
The dynamics of the process is considered as a diffusion of the DNS in the charge
and mass asymmetry coordinates, which are defined by the charge and neutron
numbers $Z$ and $N$, respectively, of the light or heavy nucleus of the DNS.
During the evolution in $Z$ and $N$, the excited DNS can decay into two fragments
by the diffusion in relative distance $R$ between the centers of the DNS nuclei.

In our approach, the production of exotic nuclei is considered as a three-step
process.
First, the initial DNS with the nucleus of ($Z=Z^{(i)},N=N^{(i)}$) is formed in
the entrance reaction channel.
And then the DNS with exotic nucleus of ($Z=Z^{(ex)},N=N^{(ex)}$) is created
through nucleon transfers.
Finally, this DNS is rapidly divided into two fragments.
The created exotic neutron-rich nuclei with ($Z^{(ex)},N ^{(ex)}$) are
predominantly de-excited $\gamma$ emissions and neutron evaporations.

In the present work, we firstly consider the reactions leading to the neutron-rich
nuclei whose excitation energies are smaller than their neutron separation energies.
Then we study the opportunity of feeding the yield of neutron-rich nuclei from
the neutron emission of properly excited more neutron-rich isotopes.

\subsection{Potential energy of dinuclear system}

In the DNS, the potential energy is defined as the sum of the mass excesses of
the both nuclei and their interaction energy.
Therefore, the phenomenological potential of the DNS is written
as~\cite{ACNPV93-ACNPV95,AAS97-AASV97a-AASV97b,AAJIM96}:
\begin{equation}
U(R, Z, N, J) = B_{1} + B_{2} + V(R, Z, N, J),
\label{eq1}
\end{equation}
where $B_{1}$ and $B_{2}$ are the mass excesses of the light and heavy nuclei,
respectively, and $V$ is the nucleus-nucleus interaction potential.
In this work, the experimental mass excess is used, if possible.
For the nuclei whose empirical mass excesses are not available, we take the
values predicted in Ref.~\cite{MNMS93} within the microscopic-macroscopic model.

The nucleus-nucleus interaction potential between two nuclei can be written as
\begin{eqnarray}
V(R, Z, N, J) &=& V_{C}(R, Z, N) + V_{N}(R, Z, N)
\nonumber \\ && \mbox{}
+ V_{\rm rot}(R, Z, N, J),
\label{eq2}
\end{eqnarray}
which includes the Coulomb interaction, the nuclear interaction, and the
centrifugal term.
The centrifugal potential $V_{\rm rot}$ in the formed DNS is usually expressed
under the assumption that this system has the rigid body moment of inertia
$\mathcal{I}$, which leads to
\begin{eqnarray}
V_{\rm rot} = \dfrac{\hbar^{2}\;J(J+1)}{2\mathcal{I}},
\label{eq3}
\end{eqnarray}
where $J$ is the orbital angular momentum of the collision between the two nuclei.

In the calculation of the Coulomb and nuclear interactions, the deformations of
the DNS nuclei are important.
The Coulomb potential for two quadrupole deformed nuclei is analytically
calculated following Ref.~\cite{Wong73} as
\begin{eqnarray}
V_{C} &=& \dfrac{e^{2}Z_{1}Z_{2}}{R} \nonumber\\
&& \mbox{} + \dfrac{e^{2}Z_{1}Z_{2}}{R^3}
\left(\dfrac{9}{20\pi} \right)^{1/2} \sum_{j=1,2} R_{0j}^2\beta_{2j}^{}
P_{2}^{} (\cos \theta_{j}^{}).
\label{eq4}
\end{eqnarray}
Here, $Z_{j}$ and $\beta_{2j}$ are the proton number and the quadrupole
deformation parameter of the nucleus ``$j$" ($j$= 1,2) in the DNS, respectively,
and $P_{2}(\cos \theta_{j})$ is the Legendre polynomial with the angle
$\theta_{j}$ between the symmetry axis of the nucleus ``$j$" and the axis $z$
connecting the centers of mass of the DNS nuclei.
The radius of a nucleus ``$j$" with mass number $A_{j}=Z_j+N_j$ is parameterized
as $R_{0j} = r_{0}A_{j}^{1/3}$, where $r_{0} = 1.15~\mbox{fm}$.
In the present work, we take the $\beta_{2j}$ values from Ref.~\cite{RNT01}.

The nuclear part of the nucleus-nucleus potential reads
\begin{eqnarray}
V_{N}  = \int \rho_{1}^{} (\bm{r}_{1}^{})
F(\bm{r}_{1}^{} - \bm{r}_{2}^{}) \rho_{2}^{} (\bm{R} - \bm{r}_{2}) \,
d \bm{r}_{1}^{} d\bm{r}_{2}^{},
\label{eq5}
\end{eqnarray}
which is obtained by using the double folding formalism~\cite{AAJIM96} with the
density-dependent nucleon-nucleon interaction.
Among those kinds of interactions, we employ the Skyrme-type effective interaction
which is well known from the theory of finite Fermi systems~\cite{Migdal}:
\begin{eqnarray}
 F(\bm{r}_{1}^{} - \bm{r}_{2}^{}) &=& C_{0} \left[ F_{\rm in}
\dfrac{\rho_0^{} (\bm{r}_{1}^{})} {\rho_{00}^{}} +
F_{\rm ex}\bigg(1-\dfrac{\rho_{0}^{} (\bm{r}_{1}^{})}{\rho_{00}^{} }\bigg)\right]
\nonumber \\ && \mbox{} \times
\delta(\bm{r}_{1}^{} - \bm{r}_{2}^{}),
\nonumber \\
F_{\rm in, ex} &=& f_{\rm in, ex} + f_{\rm in, ex}' \dfrac{(N_{1} - Z_{1})}{A_{1}}
\dfrac{(N_{2} - Z_{2})}{A_{2}},
\nonumber \\
\rho_{0}^{} (\bm{r}) &=& \rho_{1}^{} (\bm{r}) + \rho_{2}^{} (\bm{r}).
\label{eq6}
\end{eqnarray}
For our numerical calculations we use the following parameters:
$C_{0} = 300~\mbox{MeV fm}^{3}$, $f_{\rm in} = 0.09$, $f_{\rm ex} = 2.59$,
$f'_{\rm in} = 0.42$, $f'_{\rm ex} = 0.54$, and $\rho_{00} = 0.17~\mbox{fm}^{-3}$.
The densities of the nuclei are taken in the two-parameter Saxon-Woods form as
\begin{eqnarray}
\rho_{j}^{} (\bm{r}) =
\dfrac{\rho_{00}^{}}{1 + \exp \left[(r - R_{j}(\theta'_{j}, \varphi'_{j}))/a_{0j}^{} \right]}
\label{eq8}
\end{eqnarray}
with the diffuseness parameter $a_{0j}$ = 0.55~fm for the nuclei considered in this work and
$R_{j}(\theta'_{j},\varphi^{\prime}_{j})=R_{0j}[1+\beta_{2j}Y_{20}(\theta'_{j},\varphi'_{j})]$,
where $\theta^{\prime}_{j}$ and $\varphi^{\prime}_{j}$ are the angles defined in
the body-fixed system.

\subsection{Production cross sections}

The production cross section $\sigma_{Z,N}^{}$ of the nucleus with $Z$ proton
and $N$ neutrons can be written as
\begin{eqnarray}
\sigma_{Z,N}^{} &=& \sigma_{\rm cap}^{} Y_{Z,N},
\label{eq10}
\end{eqnarray}
where $\sigma_{\rm cap}^{}$ and $Y_{Z, N}$ are the capture cross section and the
formation-decay probability of the DNS configuration with the given charge and
mass asymmetries, respectively.
The capture cross section is expressed as
\begin{eqnarray}
\sigma_{\rm cap}^{} &=&
\dfrac{\pi\hbar^2}{2\mu E_{\rm c.m.}} J_{\rm cap} (J_{\rm cap}+1),
\label{eq11}
\end{eqnarray}
where $\mu$ is the reduced mass for the projectile-target system and $E_{\rm c.m.}$
is the bombarding energy in the center of mass frame.
The value of angular momentum $J_{\rm cap}$ is determined by the bombarding energy.
For example, if $E_{\rm c.m.}$ is 20 $\%$ larger than the Coulomb barrier energy,
the orbital angular momentum is about 90~\cite{ANAJ94}.
If the value of $E_{\rm c.m.}$ is close to the Coulomb barrier energy, $J_{\rm cap}$
does not exceed 30--40.
However, the angular momentum decreases the stability of the exotic nuclei of interest.
So, we choose $J_{\rm cap} = 30$ to produce neutron-rich isotopes of interest with
rather small angular momenta.

The formation-decay probability is calculated by the simple statistical method
introduced in Ref.~\cite{PAA05-PAA06}, which reads
\begin{widetext}
\begin{eqnarray}
Y_{Z,N} &\approx& 0.5\exp \left[-
\dfrac{ U \bm{(} R_{m}(Z,N_{0}),Z,N_{0},J \bm{)}
- U \bm{(}R_{m}(Z^{(i)},N^{(i)}),Z^{(i)},N^{(i)},J \bm{)} - B_{\rm qf}(Z^{(i)},N^{(i)})}
{\varTheta(Z^{(i)},N^{(i)})}
 - \dfrac{B_{R}(Z,N)}{\varTheta(Z,N_{0})}\right].
\label{eq14}
\end{eqnarray}
\end{widetext}
Here, we use the DNS potential energy $U$ at the touching distance
$R_{m}(Z,N)\approx R_{01}\Big(1 +\sqrt{5/(4\pi)}\beta_{21}\Big)
+ R_{02}\Big(1 +\sqrt{5/(4\pi)}\beta_{22}\Big)+0.5~\mbox{fm}$.
The decaying DNS with given $Z$ and $N$ has to escape from the local minimum at
$R=R_m(Z,N)$ by overcoming the potential barrier at $R=R_b(Z,N)$, where
$R_{b}(Z,N) \approx R_{m}(Z,N)+2~\mbox{fm}$.
The height of the barrier which the DNS with $Z$ and $N_{0}$ should overcome to
observe the nucleus with $Z$ and $N$ is given by
$B_{R}(Z,N) = U \bm{(} R_{b}(Z,N),Z,N,J \bm{)} - U \bm{(} R_{m}(Z,N_{0}),Z,N_{0},J \bm{)}$.
The neutron number $N_0$ corresponds to the $N/Z$ equilibrium in the DNS at
given $Z$, i.e., the conditional minimum of the potential energy surface.
In Eq.~(\ref{eq14}), the value of $B_{\rm qf}=\min(B_\eta^{\rm sym}, B_{\rm qf}^R)$
is the barrier preventing the DNS decay either in $R$ or toward more symmetric configurations.
The DNS symmetrization increases the Coulomb repulsion and, thus, leads to its
decay from more symmetric configuration.
The value of $B_{\eta}^{\rm sym}$ measures the barrier for the initial DNS toward
more symmetric configurations. The temperature $\varTheta(Z^{(i)},N^{(i)})$
is estimated using the Fermi-gas expression $\varTheta = \sqrt{E^{*}/a}$ with
the excitation energy $E^{*}(Z^{(i)},N^{(i)})$ of the initial DNS and with the
level-density parameter $a = A_{\rm tot}/$12~MeV$^{-1}$, where
$A_{\rm tot}=A_1+A_2$ is the total mass number of the DNS.
The temperature ${\varTheta(Z,N_{0})}$ is calculated for the excitation energy
$E^{*}(Z^{(i)},N^{(i)})-\bigl[ U \bm{(} R_{m}(Z,N_{0}),Z,N_{0},J \bm{)} -
U \bm{(}R_{m}(Z^{(i)},N^{(i)}),Z^{(i)},N^{(i)},J \bm{)} \bigr]$.

To reduce the number of transfered nucleons and, thus, to increase the
 yields,
 we consider only the production of neutron-rich nuclei with $Z=Z^{(i)}$
 and $N>N^{(i)}$.
 In the reactions treated the initial DNS are rather close in energy to the
 $N/Z$ equilibrium.
 In this case $\varTheta(Z^{(i)},N^{(i)})\approx \varTheta(Z,N_0)$ and
 the expression for the formation-decay probability $Y_{Z,N}$ in Eq.~\eqref{eq14}
 is simplified as
\begin{eqnarray}
Y_{Z^{(i)},N}&\approx& 0.5 \exp \left[ - \dfrac{B_{R}(Z^{(i)},N) - B_{\rm qf}(Z^{(i)},N^{(i)})}{
\varTheta(Z^{(i)},N^{(i)})} \right].
\nonumber \\
\label{eq15}
\end{eqnarray}

\section{Calculated results} \label{sec:results}

In the present work, we focus on the multi-nucleon transfers that transform the
initial DNS to a larger $N$ when $Z=Z^{(i)}$.
In order to supply a larger yield of the neutron-rich nucleus with ($Z^{(i)}$,$N$),
the potential energy of the DNS containing this nucleus should be close to that
of the initial DNS with ($Z^{(i)}$,$N^{(i)}$).
This condition should be one of the main criteria for selecting the projectile
and target nuclei.
The excitation energy of the initial DNS should not exceed the threshold above
which the excitation energy of the neutron-rich product is larger than its
neutron separation energy.
So, the nuclei produced near the neutron drip line should be quite cold to
avoid their losses during the de-excitation process.
The primary neutron-rich nucleus excited above the threshold contributes to the
yield of the nucleus with smaller number of neutrons.
Because the exotic nucleus results from the multi-nucleon transfers between the
projectile-like and target-like parts of the DNS, one can assume the thermal
equilibrium in the DNS containing the exotic nucleus or in the DNS which is quite
far from the initial DNS in the $(Z,N)$ space.
Indeed, for the formation of these DNS, quite a long time is needed such as
$t_{0} \approx 10^{-20}~\mbox{s}$ at $J \leq 30$.
Therefore, in the considered reactions the excitation energy of the light or
heavy nucleus with the mass $A_{1}$ or $A_{2}$ is
\begin{eqnarray}
E_{1,2}^{\ast}(Z^{(i)},N) = \left[E^{\ast}(Z^{(i)},N_{0}) - B_{R}(Z^{(i)},N)\right]
\dfrac{A_{1,2}}{A_{\rm tot}}.
\nonumber \\
\label{eq1e}
\end{eqnarray}
The deviation from the thermal equilibrium is expected only for the DNS near the
initial DNS with $(Z^{(i)},N^{(i)})$ which are formed in a short time.
If one of the DNS nucleus is magic, it takes a smaller excitation energy than the
value determined by Eq.~(\ref{eq1e}).
So, our estimates of production cross sections with Eq.~(\ref{eq1e}) indicate
the upper limits of the yields.
The cross section $\sigma_{Z^{(i)},N}$ for the production of the exotic nucleus
with $(Z^{(i)},N)$ increases with $E^{\ast}(Z^{(i)},N_{0})$ until when
$E_{1}^{\ast}(Z^{(i)},N)$ or $E_{2}^{\ast}(Z^{(i)},N)$ becomes equal to the neutron
separation energy $S_{n}(Z^{(i)},N)$.
Up to this moment the primary and secondary yields coincide.
A further increase of $E^{\ast}(Z^{(i)},N_{0})$ would increase the yield of the
primary excited neutron-rich nuclei which then emit neutrons contributing to the
yield of secondary nuclei with smaller number of neutrons.
The mass excesses and, correspondingly, the neutron separation energies $S_{n}$
for unknown nuclei are taken from the predictions of the finite-range
liquid-drop model presented in Ref.~\cite{MNMS93}.

Through Refs.~\cite{PAA05-PAA06,AAL10,AAS03,KAAS10-KAAS11-KAASW11} the
applicability of the present model to describe the existing experimental data
has been tested.
The yields of isotopes, which are several nucleons away from the entrance channel,
are well described by the DNS potential energy, and our approach therefore seems
to be suitable.

\begin{figure}[t] \centering
\includegraphics[width=1.15\columnwidth,angle=0,clip]{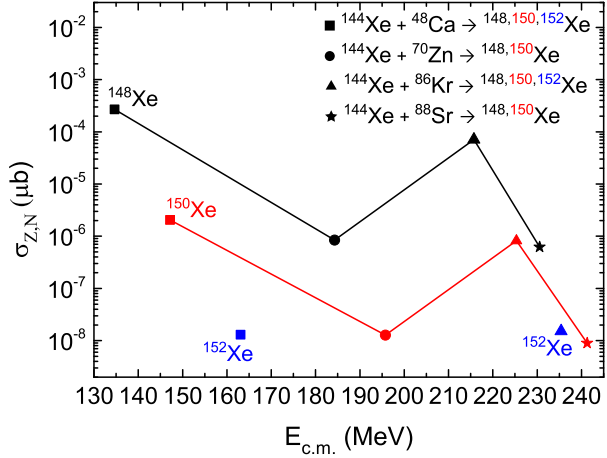}
\caption{\label{fig:1} (Color online)
The predicted cross sections of the production of neutron-rich isotopes
\nuclide[148,150,152][]{Xe} in the transfer reactions with the radioactive beam
of \nuclide[144][]{Xe}.
The cross sections correspond to the maxima of $0n$ evaporation channels.
The symbols correspond to the indicated reactions.}
\end{figure}

Employing the radioactive beam of \nuclide[144][]{Xe}, one can produce more
neutron-rich isotopes of \nuclide[][]{Xe} and study their properties.
The nuclei \nuclide[148,150,152][]{Xe} have more than 93 neutrons, i.e. with
these nuclei one can trace an evolution of neutron shells well beyond $N=82$.
The predicted cross sections in $0n$ evaporation channel are presented in
Fig.~\ref{fig:1}.
Here, we show only the cross sections larger than $0.01~\mbox{pb}$.
In the reactions considered in this work, the formation of the nuclei of interest
occurs in accordance with the following schemes:
$\nuclide[144][]{Xe} + \nuclide[48][]{Ca} \to \nuclide[142][]{Xe} + \nuclide[50][]{Ca}
\to \nuclide[148,150,152][]{Xe} + \nuclide[44,42,40][]{Ca}$,
$\nuclide[144][]{Xe} +\nuclide[70][]{Zn} \to \nuclide[136][]{Xe} + \nuclide[78][]{Zn}
\to \nuclide[148,150][]{Xe} + \nuclide[66,64][]{Zn}$,
$\nuclide[144][]{Xe} + \nuclide[86][]{Kr} \to \nuclide[138][]{Xe} + \nuclide[92][]{Kr}
\to \nuclide[148,150,152][]{Xe} + \nuclide[82,80,78][]{Kr}$,
$\nuclide[144][]{Xe} + \nuclide[88][]{Sr} \to \nuclide[136][]{Xe} + \nuclide[96][]{Sr}
\to \nuclide[148][]{Xe} + \nuclide[84][]{Sr}$.
The \nuclide[48][]{Ca} target seems to be favorable to produce
\nuclide[148,150,152][]{Xe}, while the \nuclide[86][]{Kr} target would lead to
relatively large cross sections for the yields of \nuclide[148,150][]{Xe}.
With increasing charge number of the target nucleus the initial DNS becomes
unstable with respect to the decay in $R$.
So, the use of the combination with a larger $Z_1 \times Z_2$ does not seem to
be promising even they supply a gain in $Q$ values.

\begin{figure}[t] \centering
\includegraphics[width=1.15\columnwidth,angle=0,clip] {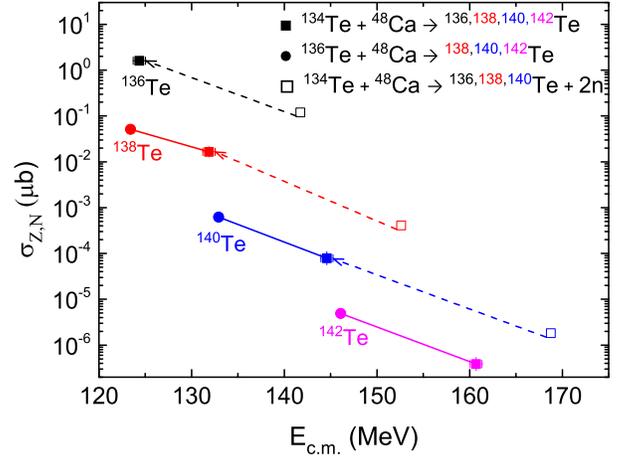}
\caption{\label{fig:2} (Color online)
The predicted cross sections of the production of neutron-rich isotopes
\nuclide[136,138,140,142][]{Te} in the transfer reactions
$\nuclide[134,136][]{Te} + \nuclide[48][]{Ca}$.
The cross sections correspond to the maxima of the $0n$ evaporation channels
(closed symbols) and $2n$ evaporation channels (open symbols).
The symbols correspond to the indicated reactions.
Arrows connect the points corresponding to the productions of the same isotope
but in different evaporation channels.}
\end{figure}

Figure~\ref{fig:2} shows the cross sections predicted for the production of
neutron-rich isotopes of \nuclide[][]{Te} with the beams of \nuclide[134,136][]{Te}.
Only the yields with the cross sections larger than $0.1~\mbox{pb}$ are presented.
As in the case of \nuclide[][]{Xe} beams, the \nuclide[48][]{Ca} target is
expected to be the most appropriate to create Te isotopes close to the neutron
drip line.

For these calculations we use the neutron separation energies $S_n$ predicted in
Ref.~\cite{MNMS93}.
As a result, the model-dependence of these values is unavoidable.
To study the influence of the uncertainties of $S_n=4\mbox{---}4.5~\mbox{MeV}$,
in the case of the $\nuclide[136][]{Te} + \nuclide[48][]{Ca}$ reaction,
we consider the cases of $S_n \pm 0.5~\mbox{MeV}$.
The variations of the obtained cross sections are marked by the error bars in
Fig.~\ref{fig:2}.
We find that 0.5~MeV uncertainty in $S_n$ causes about 20\% uncertainty in the
cross section and about 1~MeV uncertainty in the corresponding value of
$E_{\rm c.m.}$.

If the bombarding energy is higher, then the neutron evaporation process occurs
from the primary excited products.
Here, we also consider the $2n$ evaporation channels as in the reactions of
$ \nuclide[134][]{Te} + \nuclide[48][]{Ca} \to \nuclide[136,138,149][]{Te} + 2n$.
As shown in Fig.~\ref{fig:2}, the yields of the primary products with the same
$N$ are larger in the $2n$ channel than in the $0n$ channel because of the
larger excitation.
After the 2-neutron emission, the secondary products ($N-2$) appear with the
same cross sections as the primary products ($N$) which are almost one order of
magnitude smaller than those for the products produced in the $0n$ evaporation
channel.
So, the increase of $E_{\rm c.m.}$ feeds the yield of isotopes with smaller
number of neutrons.
Although this feeding is quite small for the isotopes considered in this present
work, one may increase the thickness of the target in future experiments in
order to increase the total yields of the isotopes of interest.
We note that the role of the feeding becomes more important as decreasing the
neutron number of the produced isotope.

\begin{figure}[t] \centering
\includegraphics[width=1.15\columnwidth,angle=0,clip]{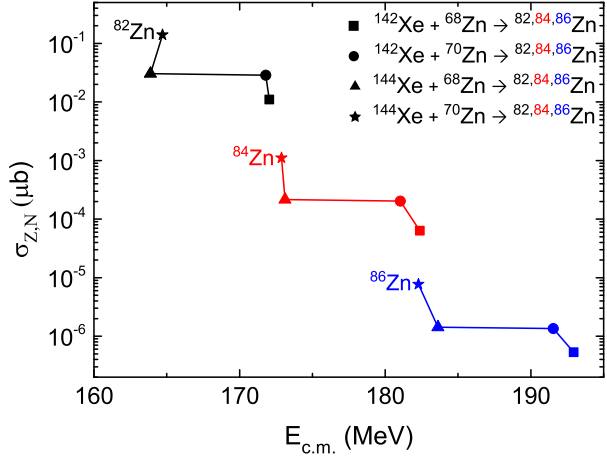}
\caption{\label{fig:3}  (Color online)
The expected maximal cross sections of the production of neutron-rich isotopes
\nuclide[82,84,86][]{Zn} in the $0n$ evaporation channels of the transfer reactions
$\nuclide[142,144][]{Xe} + \nuclide[68,70][]{Zn}$.
The symbols correspond to the indicated reactions.}
\end{figure}

With the radioactive beams of \nuclide[142,144][]{Xe} one can produce
neutron-rich target-like nuclei with $N>50$.
In Fig.~\ref{fig:3} the cross sections are predicted for the reactions of
$\nuclide[142,144][]{Xe} + \nuclide[68,70][]{Zn} \to \nuclide[82,84,86][]{Zn}$.
The yields of the neutron-rich \nuclide{Zn} isotopes increase with neutron numbers
of the projectile and target.
The yield of neutron-rich isotopes decreases almost by two orders of magnitude
when its neutron number increases by 2 units.
The reactions of $\nuclide[142][]{Xe} + \nuclide[70][]{Zn}$ and
$\nuclide[144][]{Xe} + \nuclide[68][]{Zn}$ are found to have the same value of
the cross sections but different values of the optimal $E_{\rm c.m.}$.
The production of \nuclide[86][]{Zn} occurs with the largest cross section in
the $\nuclide[144][]{Xe} + \nuclide[70][]{Zn}$ reaction.
Therefore, the shell evolution beyond $N=50$ can be traced by studying
neutron-rich nuclei in this region of $Z$.

\begin{figure}[t] \centering
\includegraphics[width=1.15\columnwidth,angle=0,clip]{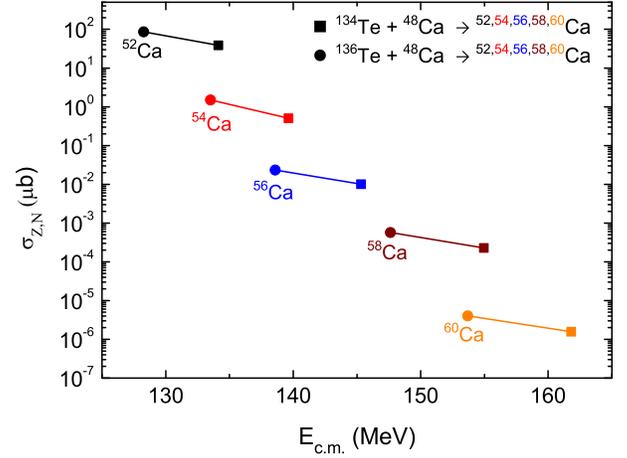}
\caption{\label{fig:4}  (Color online)
The expected maximal cross sections of the production of neutron-rich isotopes
\nuclide[52,54,56,58,60][]{Ca} in the $0n$ evaporation channels of the transfer
reactions $\nuclide[134,136][]{Te} + \nuclide[48][]{Ca}$.
The symbols correspond to the indicated reactions.}
\end{figure}

The reactions of $\nuclide[134,136][]{Te} + \nuclide[48][]{Ca}$ can be used to
produce the neutron-rich isotopes \nuclide[52,54,56,58,60][]{Ca} ($N>28$) with
the cross sections larger than 1~pb as shown in Fig.~\ref{fig:4}.
The replacement of \nuclide[134][]{Te} by \nuclide[136][]{Te} leads to about
5 times larger yields of neutron-rich \nuclide[][]{Ca} isotopes.
However, these cross sections are about 5 times smaller compared with the
$\nuclide[132][]{Sn} + \nuclide[48][]{Ca}$ reaction~\cite{AAL10}.

\begin{figure}[t]\centering
\includegraphics[width=1.15\columnwidth,angle=0,clip]{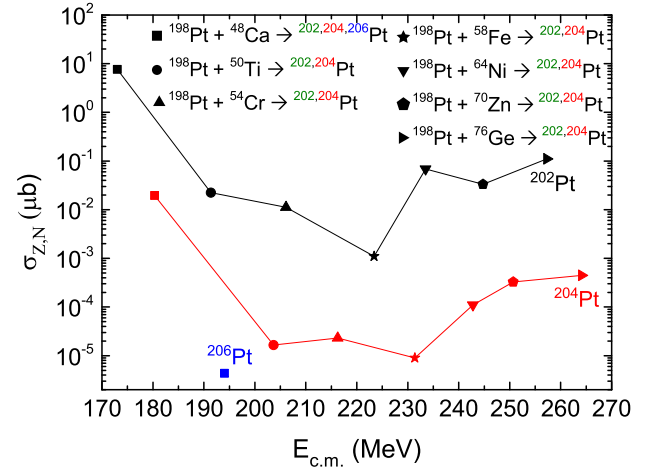}
\caption{\label{fig:5}  (Color online)
The expected maximal cross sections of the production of neutron-rich isotopes
\nuclide[202,204,206][]{Pt} in the $0n$ evaporation channels of the listed
transfer reactions.
The symbols correspond to the indicated reactions.}
\end{figure}

To study the neutron-rich nuclei with $N$ around 126, we consider the production
of \nuclide[202,204,206][]{Pt}.
The unknown platinum isotopes can be produced in the nucleon transfer reactions
of $\nuclide[198][]{Pt} + \nuclide[48][]{Ca}, \nuclide[50][]{Ti}, \nuclide[54][]{Cr},
\nuclide[58][]{Fe}, \nuclide[64][]{Ni}, \nuclide[70][]{Zn}, \nuclide[76][]{Ge}
\to \nuclide[202,204,206][]{Pt}$ with stable beams.
The estimated cross sections are presented in Fig.~\ref{fig:5}.
The $\nuclide[198][]{Pt} + \nuclide[48][]{Ca}$ reaction is the most favorable
to produce the \nuclide[202,204,206][]{Pt} isotopes.
For the projectiles with larger value of $Z$ the cross sections are about two
orders of magnitude smaller.
In the reactions with \nuclide[64][]{Ni}, \nuclide[70][]{Zn}, and
\nuclide[76][]{Ge} the cross sections are similar to or larger than those of the
reactions with \nuclide[50][]{Ti}, \nuclide[54][]{Cr}, and \nuclide[58][]{Fe} due
to the gains in the $Q$-values.
With the reaction partner heavier than \nuclide[76][]{Ge} the DNS formed in this
case becomes very unstable with respect to the decay in $R$ and, therefore, the
nucleon transfers are reduced.
This shows that the use of more symmetric reactions does not help to produce
isotopes close to the neutron drip line.

\begin{figure}[t] \centering
\includegraphics[width=1.15\columnwidth,angle=0,clip]{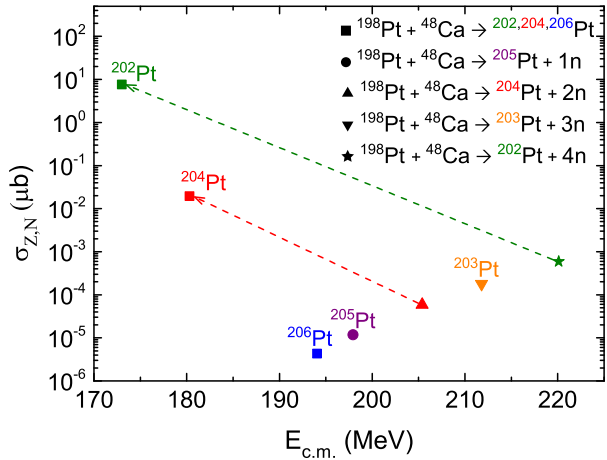}
\caption{\label{fig:6}  (Color online)
The expected maximal cross sections of production of the neutron-rich isotopes
\nuclide[202,204,206][]{Pt} in the $(0 \sim 4)n$ evaporation channels of the
transfer reactions of $\nuclide[198][]{Pt} + \nuclide[48][]{Ca}$.
The symbols correspond to the indicated reactions.
Arrows connect the points corresponding to the productions of the same isotope
in different evaporation channels.}
\end{figure}

In Fig.~\ref{fig:6} we compare the cross sections in different neutron
evaporation channels.
Although the cross section of the production of \nuclide[204][]{Pt} in the $2n$
evaporation channel is more than two orders of magnitude smaller than that of
the $0n$ evaporation channel, it is larger than 0.1~nb and could be examined
with the current experimental facilities.
In order to make these experiments feasible, thicker targets and higher energy
will be helpful as they would increase the yield.

\begin{figure}[t] \centering
\includegraphics[width=1.15\columnwidth,angle=0,clip]{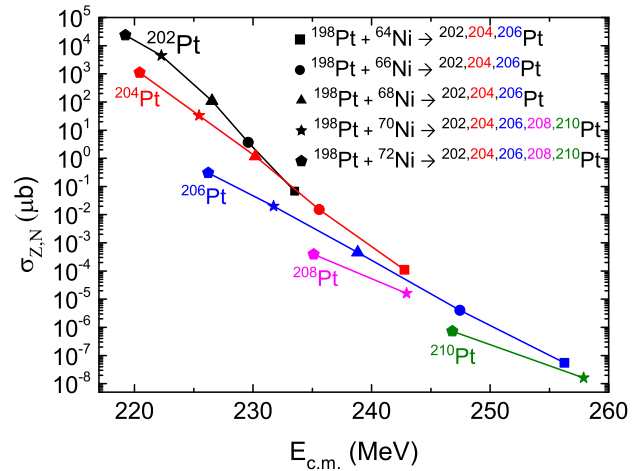}
\caption{\label{fig:7}  (Color online)
The expected maximal cross sections of production of the neutron-rich isotopes
\nuclide[202,204,206,208,210][]{Pt} in the $0n$ evaporation channels of the
transfer reactions of $\nuclide[198][]{Pt} + \nuclide[64,66,68,70,72][]{Ni}$.
The symbols correspond to the indicated reactions.}
\end{figure}

The results shown in Fig.~\ref{fig:7} show that the use of the radioactive beams
of \nuclide[66,68,70,72][]{Ni} and the \nuclide[198][]{Pt} target will allow to
increase the yields of the neutron-rich nuclei \nuclide[202-210][]{Pt}
in comparison with the case of the \nuclide[64][]{Ni} beam.
The increase of the mass number of \nuclide{Ni} by 2 units leads to more than
10 times larger cross sections.
In the $\nuclide[72][]{Ni} + \nuclide[198][]{Pt}$ reaction one can produce even
$^{210}$Pt isotope ($N$=132) with the cross section of about 1~pb.

\section{Summary} \label{sec:summary}

In this work, we propose the multi-nucleon transfer reactions with radioactive
and stable beams to produce the neutron-rich isotopes close to the neutron
drip-line.
The radioactive beams are expected to be of a good intensity enough to produce
such isotopes properly even with a small cross section.
Thus we search for the production of neutron-rich isotopes of which cross
sections are larger than 1~pb as the experiments for such reactions can be
performed at current and future facilities.
In the reactions considered in the present work, the neutron-rich nuclei with
neutron numbers beyond the magic values
$N=28$ (\nuclide[52,54,56,58,60][]{Ca}), $N=50$ (\nuclide[82,84,86][]{Zn}),
$N=82$ (\nuclide[136,138,140,142][]{Te} and \nuclide[148,150,152][]{Xe}),
and $N=126$ ($^{202,204,206,208}$Pt) would be produced with the cross sections
larger than 1~pb.
The study of these nuclei is, therefore, desirable to trace the shell evolution
toward the neutron drip line.
The results of the present calculations show that the production cross sections
of neutron-rich nuclei increase by about one order of magnitude with increasing
mass number by 2 units in the reaction partners.
The maximal cross sections are found to be in the $0n$ evaporation channels.
The cross sections in $1n \sim 4n$ evaporation channels are smaller but still
measurable.
Then in this case, one can use higher bombarding energies and thicker targets to
increase the yields of neutron-rich nuclei.
Some of the reactions investigated in this work can be tested by the current
experimental facilities.
If the expected intensity of the radioactive beam is $10^{7}-10^{9}$
 pps, then the production rate of the nuclides
 is roughly estimated as 0.01--1 particle per day at production cross
 section 1 nb.
 Therefore, the observation of the isotopes produced with the cross section
 smaller than 1 nb would be
 difficult in the experiments proposed with the present radioactive beams.
 The production cross
 section of 1 pb would lead to the reasonable irradiation time if the
 intensity of the radioactive beam
 is at least larger than $10^{10}$ pps.

\acknowledgments
The work of Y.K. and M.-H.M. was supported by the Rare Isotope Science Project
of Institute for Basic Science funded by Ministry of Science, ICT and Future
Planning and National Research Foundation of Korea (2013M7A1A1075766).
The work of G.G.A. and N.V.A. was supported in part by the RFBR.
Y.O. was supported in part by the National Research Foundation of Korea under
Grant No.~NRF-2013R1A1A2A10007294.

\end{document}